# When Support Hides Progress: Insights from a Physics Tutorial on Solving Laplace's Equation Using Separation of Variables in Cartesian Coordinates

Jaya Shivangani Kashyap *, Robert Devaty and Chandralekha Singh

Department of Physics and Astronomy, University of Pittsburgh, Pittsburgh, PA, 15260 USA; devaty@pitt.edu (R.D.); clsingh@pitt.edu (C.S.)
* Correspondence: jak509@pitt.edu

**Abstract**

The electrostatic potential in certain types of boundary value problems can be found by solving Laplace's Equation (LE). It is important for students to develop the ability to recognize the utility of LE and apply the method to solve physics problems. To develop students' problem-solving skills for solving problems that can be solved effectively using Laplace's equation in an upper-level electricity and magnetism course, we developed and validated a tutorial focused on finding electrostatic potential in a Cartesian coordinate system. The tutorial was implemented across three instructors' classes, accompanied by scaffolded pretest (after traditional lecture) and posttest (after the tutorial). We also conducted think-aloud interviews with advanced students using both unscaffolded and scaffolded versions of the pretest and posttest. Findings reveal common student difficulties that were included in the tutorial as a guide to help address them. The difference in the performance of students from the pretest after lecture to the posttest after the tutorial was similar on the scaffolded version of the tests (in which the problems posed were broken into sub-problems) for all three instructors' classes and interviewed students. Equally importantly, interviewed students demonstrated greater differences in scores from the pretest and posttest on the unscaffolded versions in which the problems were not broken into sub-problems, suggesting that the scaffolded version of the tests may have obscured evidence of actual learning from the tutorial. While a scaffolded test is typically intended to guide students through complex reasoning by breaking a problem into sub-problems and offering structured support, it can limit opportunities to demonstrate independent problem-solving and evidence of learning from the tutorial. Additionally, one instructor's class underperformed relative to others even on the pretest. This instructor had mentioned that the tests and tutorial were not relevant to their current course syllabus and offered a small amount of extra credit for attempting to help education researchers, highlighting how this type of instructor framing of instructional tasks can negatively impact student engagement and performance. Overall, in addition to identifying student difficulties and demonstrating how the tutorial addresses them, this study reveals two unanticipated but critical insights: first, breaking problems into sub-parts can obscure evidence of students' ability to independently solve problems, and second, instructor framing can significantly influence student engagement and performance.

**Keywords:** Tutorial, Electrostatics, Laplace's Equation, Physics, Scaffolding, Problem-solving, Instructor framing, Research-based Learning Tools, Incentive to Learn

## 1. Introduction

Electricity and Magnetism (E&M) is recognized as a conceptually and mathematically intensive subject, which is challenging for students at both the introductory and advanced levels [1]. Our investigation described here seeks to contribute to this body of research by developing and evaluating a research-based tutorial designed to support student learning of boundary value problems involving Laplace's equation. The goal of this research-based tutorial on Laplace's equation is to provide scaffolding in the context of an upper-level E&M course. We identified common student difficulties with physics problems that can be solved effectively using Laplace's equation and incorporated them in the tutorial to improve students' understanding and their problem-solving ability. The sensemaking of some of the advanced students observed during individual think-aloud interviews conducted as a part of development and validation of the tutorial is discussed in [2, 3]. The tutorial was implemented along with the scaffolded versions (problems broken down into sub-problems) of the pretest and posttest across three different instructors' classes at a large research university in the United States. The pretest was administered in class after students had received traditional instruction on the topic, then students were given the tutorial as homework followed by an in-class posttest.

We note that in individual interviews we conducted with advanced students, they were also asked to solve the unscaffolded version of the pretest (and posttest) followed by the scaffolded version of each. While our initial focus was on evaluating learning outcomes and refining the tutorial based on observed student struggles and minimizing expert blind spots that crop up in learning tools developed only based upon cognitive task analysis from an expert perspective, our analysis revealed two additional insights that were not anticipated at the outset. First, the use of both scaffolded and unscaffolded assessments in the interviews (but only the scaffolded version in the in-class implementation of the tutorials) helped us explore how scaffolded tests can somewhat obscure evidence of student learning from the tutorial. Second, when one of the instructors mentioned to students that the tests and the tutorial were part of a research study and not relevant to the current course syllabus and offered a very small amount of extra credit for attempting rather than for correctness on the tests and tutorial, student engagement with the material may have been diminished since their performance was noticeably worse.

The rest of the sections are organized as follows. We first conduct a literature review summarizing research literature on integration of mathematics and physics in physics problem-solving since such integration plays a central role in student learning of E&M not only at the introductory level but also at the advanced level. Then, we describe our framework and methodology before focusing on results and discussion. Finally, we focus on broader discussion, instructional implications, limitations and future work.

## 2. Literature Review and Framework

*2.1. Integration of Mathematics and Physics in Physics Problem-Solving*

Physics in general and E&M in particular demand fluency with mathematical techniques such as vector calculus, trigonometry, and differential equations, along with the ability to integrate mathematical knowledge within a physics context. Prior work has documented challenges with such integration in physics in general and E&M in particular [4-7], as well as in the context of applications of boundary conditions and symbolic reasoning [8, 9]. Other studies have focused on student use of vector calculus concepts such as divergence, curl, and graphical representations of vector fields [9-12], as well as their understanding of multivariable calculus concepts [13, 14]. Broader discussions on transfer and mathematical reasoning in physics provide further context for these investigations [15-19]. Mathematics in physics problem-solving is not just a computational aid; it plays a foundational role in how students interpret and understand physical systems to solve problems [20-26].

Research further shows that mathematical and physical knowledge are deeply intertwined, with students' epistemological framing influencing how they blend these domains for problem-solving [27-30]. Students often encounter challenges in physics because they are still developing expertise and have difficulty in integrating physics and mathematics [31-36]. As a result, they may not consistently draw upon relevant mathematical knowledge from prior coursework or may struggle to coordinate mathematics with physics while solving problems [37-42]. Uhden et al. [43] described two distinct roles of mathematics in physics: technical and structural, i.e., "the technical skills are associated with pure mathematical manipulations whereas the structural skills are related to the capacity of employing mathematical knowledge for structuring physical situations". This framework is consistent with the viewpoint of Tzanakis [44], "mathematics is the language of physics, not only as a tool for expressing, handling and developing logically physical concepts, methods, and theories, but also as an indispensable, formative characteristic that shapes them, by deepening, sharpening, and extending their meaning, or even endowing them with meaning". In alignment with Uhden et al. [43], we acknowledge that mathematics in physics serves both structural and technical functions. The structural role is particularly significant, as it is so deeply interwoven with the physical concepts that separating the two is often not possible.

*2.2. Prior Investigations on Student Understanding in the Context of E&M*

Given the deep interconnection between mathematics and physics, it is essential to investigate the challenges students face while learning E&M concepts and provide them with appropriate support to address these difficulties. In response, prior research has documented student difficulties with E&M concepts at the introductory level including electric and magnetic fields [45], electric field, field lines and superposition principle [46-49], electric flux [50], Gauss's law [51, 52], electromotive force and Faraday's law [53-56], Ampére-Maxwell's law [57] and issues related to electric circuits [58, 59]. Studies at introductory level have also investigated how to improve students' representational consistency in the context of Gauss's law [60], student understanding of transients in direct current electric circuits [61], the missing link between electrostatics and electrodynamics in students' reasoning [62], student understanding of conductors and insulators [63] and capacitance [64-66].

In advanced E&M, researchers have made significant efforts to strengthen the understanding of E&M among advanced students. Singh [67], for example, using a research-based instrument, investigated student understanding of symmetry and Gauss's law from introductory to graduate levels, while Bilak and Singh [68] explored introductory and graduate student understanding of insulators and conductors. Chasteen et al. [69] redesigned a junior-level E&M course and developed a diagnostic instrument to assess upper-level student understanding of electrostatics [70], which has also been used to assess student learning of these concepts at other institutions [71]. Pepper et al. [72] examined the challenges upper-division students face in applying mathematics in E&M contexts. A series of studies have documented persistent challenges. These include difficulties with boundary conditions for electromagnetic waves by Ryan et al. [8], difficulties using vector calculus in electrodynamics [10], and using divergence and curl in electromagnetism [11], along with difficulties interpreting symbolic and graphical vector field representations by Bollen et al. [9]. Bollen et al. further developed and assessed a guided inquiry-based teaching-learning sequence aimed at improving student understanding of vector calculus concepts in electrodynamics [12]. Singh and Maries [73] investigated the understanding of divergence and curl among physics graduate students. Wilcox and Corsiglia [6] researched student difficulties in integrating concepts across magnetostatics and classical mechanics contexts. Mason et al. [74] studied performance improvements in upper-division E&M when students were given explicit incentives to correct mistakes. In the upper-level electrostatics context [2, 3], case studies elaborating on individual students' sensemaking while solving problems involving method of images and Laplace's equation complement the quantitative study presented here.

*2.3. Student Understanding of Laplace's Equation*

Since a specific topic that poses significant difficulty in upper-level E&M is the application of Laplace's equation method to solve certain types of boundary value problems, we focus on this issue in this research. Wilcox and Pollock [75] analyzed student difficulties with separation of variables to solve Laplace's equation, since it is a standard approach for finding the electrostatic potential for certain type of boundary value problems. In particular, in upper-level E&M, students learn to recognize the applicability and solve Laplace's equation $\nabla^2 V = 0$ to find the electrostatic potential $V$ in the region, where the charge density is zero. First, students should be able to identify whether Laplace's equation can be used effectively to find the potential in a given configuration. Once students recognize that Laplace's equation is an appropriate method for solving the given problem, they can follow a standard approach to solve the partial differential equation by finding separable solutions and using them to find the general solution and then the unique solution of the potential in the region of interest using the remaining boundary conditions to evaluate the coefficients. As noted in the preceding section, Ref. [2] describe a case study detailing the individual sensemaking of advanced physics students while solving for electrostatic potential using Laplace's equation and is synergistic to the quantitative study presented here.

*2.4. Scaffolding While Problem-Solving and Zone of Proximal Development*

One approach to improving student learning in introductory physics commonly used in physics education, e.g., by the University of Washington group, involves research-based tutorials [59, 76-81]. Our research on tutorial development is guided by the role of scaffolding in supporting students as they work through complex physics problems [82-84]. Scaffolding in this context refers to providing tailored support that guides students through complex tasks, while gradually building the skills and knowledge needed for independent problem-solving. In physics education, scaffolding techniques include breaking a problem into smaller, manageable sub-problems, offering hints, guiding questions, or providing worked examples. The idea is that, with appropriate support, students can progress from basic skills to more advanced problem-solving. These strategies can reduce cognitive load, help focus attention on critical aspects of a problem and foster metacognitive skills [85]. This research is guided by the theoretical framework of the Zone of Proximal Development (ZPD), inspired by Vygotsky's work [86]. The ZPD refers to the gap between what a learner can accomplish independently and what they can achieve with appropriate scaffolding support from an expert or in collaboration with peers. This framework highlights that meaningful learning happens when the tasks students engage with fall within their ZPD, which itself is dynamic and expands as students advance in their level of understanding. The balance between sufficient support and helping learners develop independence in problem-solving is especially critical [60] while learning advanced physics concepts that require integration of mathematics and physics. Therefore, effective instruction should align with students' existing knowledge and build on it. Carefully designed instructional activities, such as the guided inquiry-based sequences in the tutorial on Laplace's equation, can help expand students' ZPD and foster deeper understanding while improving student problem-solving skills. In particular, scaffolding support grounded in ZPD [86] can help students learn and solve complex problems. This framework is synergistic with the cognitive apprenticeship model that emphasizes the importance of modeling to help students understand the criteria of good performance, coaching and scaffolding critical for helping students engage meaningfully in learning while receiving prompt feedback as needed, and fading of the support to ensure that students develop self-reliance [87].

*2.5. Instructor Framing and Student Engagement*

Additionally, the way an instructor frames the benefits of a learning tool, e.g., a tutorial and accompanying assessments, as well as the incentives they offer for meaningful engagement with the learning tools, can significantly influence student engagement and learning outcomes. Prior research [88] has shown that students' motivation and performance are strongly influenced by how instructional activities are framed, and whether they are presented as an integral part of the course or as activities for completion without any relevance for the course syllabus. When students perceive tasks as irrelevant or of low consequence for a course, they are less likely to invest effort, regardless of their learning potential [88].

## 3. Methodology

*3.1. Tutorial Development and Validation*

The tutorial (along with the pre-/posttests) was designed for the upper-level undergraduate E&M course. It uses a guided inquiry-based approach to learning and provides scaffolding support to help students solve Laplace's equation $\nabla^2 V = 0$ for the electrostatic potential V. The tutorial begins by helping students identify whether Laplace's equation is appropriate for a given physical configuration, based on the absence of charge in the region of interest with boundary conditions. Students are then encouraged to conceptually analyze the problem, e.g., via diagrams, and choose an appropriate coordinate system. Then, students are scaffolded to separate the partial differential equation into ordinary differential equations to obtain separable solutions, which are products of functions of a single variable. A key step is ensuring that the separation constants satisfy the condition that their sum equals zero. Students solve the resulting ordinary differential equations, apply boundary conditions to determine unknown constants, and construct the general solution using a superposition of separable solutions. Students are then guided to obtain a unique solution representing the electrostatic potential in the region of interest for the given configuration using the Fourier trick and obtain the expansion coefficients in the superposition using the remaining boundary condition(s). The guided inquiry-based tutorial is designed to support students in navigating each of these aspects of problem-solving systematically.

We used common student difficulties in solving Laplace's equation in E&M as a guide to develop the tutorial. Students' difficulties were investigated via discussions with course instructors and students as well as individual interviews with students. After developing the preliminary version of the pre-/posttests and the tutorial based upon a cognitive task analysis from an expert perspective as well as individual student feedback, the tutorial was implemented over a period of two years in the classroom of instructor 1 (I1) after traditional lecture-based instruction on relevant concepts. To improve the tutorial (along with corresponding pretest and posttest) and reduce student difficulties, we iteratively made changes based on think-aloud interviews conducted with 13 students (8 graduate and 5 undergraduate students who had lecture-based instruction in relevant concepts earlier) over the period. As we did not observe significant qualitative differences in the sensemaking of graduate and undergraduate students while problem-solving, we combined their data for the purposes of analysis. The tutorial and the corresponding pretest and posttest were also implemented in the classes of two more instructors, instructor 2 (I2) and instructor 3 (I3). It was continually revised through ongoing discussions among the three researchers (authors, two of whom have taught the upper-level E&M course), student interviews, and feedback from other physics instructors. During the validation process, we ensured content validity of the tutorial and the accompanying assessments (pretest and posttest) through iterative feedback from physics instructors experienced in teaching upper-level E&M throughout the development process.

The tutorial uses a guided inquiry-based approach to learning and has scaffolded problems with learning sequences that build on each other to help students develop expertise in finding the potential in electrostatics problems for a given configuration (when Laplace's equation is valuable for solving the problem). Along with the tutorial, the pretest and posttest on the underlying concepts that were developed and validated have similar problems with a slight modification in the boundary conditions.

*3.2. Pre-/Posttests and Learning Objectives of the Tutorial*

The pretest and posttest on Laplace's equation consist of different problems of similar difficulty levels. Two versions of each test were developed: unscaffolded and scaffolded. The unscaffolded version of the test consists of a problem statement analogous to a typical textbook style problem. Students are expected to solve the unscaffolded pre-/posttest problems without any extra support. On the other hand, the scaffolded version of the tests (the only version used in the in-class implementation) provides extra support to students by breaking down the problem into sub-problems and asking them to solve the sub-problems.

The learning objectives of the tutorial consistent with the concepts evaluated in the pretest and posttest are shown in Table 1. The unscaffolded version of the pretest and posttest evaluate student ability to identify whether solving Laplace's equation is an effective approach to find the potential and then apply the method of separation of variables to find the potential. The unscaffolded test assesses all of the learning objectives of the tutorial shown in Table 1, except the behavior of electric field lines and equipotential surfaces, explicitly. On the other hand, the scaffolded version of the tests breaks the problem into smaller components and provides targeted prompts aligned with specific learning objectives (but not all).

Table 1. Learning objectives of the tutorial and concepts evaluated in the latest version of the scaffolded pretest and posttest on Laplace's equation. Each of the learning objectives starts with "Students should be able to…".

| Learning Objectives | Pretest or Posttest Question (Latest Version) | Corresponding Tutorial Question |
|---|---|---|
| identify the situation(s) in which Laplace's equation can be used to find the potential | 1 | 1, 2 |
| identify a relevant coordinate system and coordinates | | 3, 11, 28 (d) |
| draw a diagram of the situation | | 4, 28 (a) |
| write and/or label or interpret boundary conditions for the given situation | 3 | 5, 10, 28 (b) |
| write Laplace's equation for the relevant coordinate system | | 6, 12 |
| apply/explain the method of separation of variables | 2 | 7, 8, 9, 28 (c) |
| apply the boundary conditions to find different constants | | 14, 15, 16, 18, 28 (e), 28 (f) |
| write the separable solution | 4 | 13, 17 |
| write the general solution as a linear combination of separable solutions | 5 | 19, 20, 28 (g) |
| evaluate the expansion coefficients by applying the remaining boundary condition and using the Fourier trick | 6 | 21, 22, 23, 24, 27 |
| write the unique solution using the expansion coefficients | | 25, 28 (h), 28 (i), 28 (j) |
| draw the behaviors of electric field lines and equipotential surfaces for the given problem | | 26 |

The guided inquiry-based teaching-learning sequences in the tutorial build on each other. The first part of the tutorial focuses on a two-dimensional problem, for which sufficient scaffolding is provided. In the second part, the complexity is increased as the problem becomes three-dimensional, and the level of scaffolding is gradually reduced. In the final version of the tutorial, checkpoints are provided after students work through a section so that they can reflect upon their mistakes and correct them before moving forward. The question numbers from the pretest, posttest, and tutorial shown in Table 1 are from the latest versions, they might differ for some students who were interviewed earlier. In this paper, questions from the scaffolded version of the pretest and posttest are simply referred to as scaffolded pretest Q1, scaffolded posttest Q1, and so on.

*3.3. Interview Details*

Discussions with students early in the development and validation of the tutorial clarified their difficulties and what approaches may help in improving their understanding. We also interviewed 13 advanced students at various stages using a think-aloud protocol. In the temporal order, the think-aloud interviews were conducted with a cluster of six physics graduate students who had taken an upper-level E&M course earlier, a second cluster of five upper-level physics undergraduates who were either currently enrolled (and had instruction in relevant concepts) or had taken this course previously, and finally with two additional physics graduate students. These interviews were recorded on Zoom and students used a device to scribble on the screen when they solved the problems while thinking aloud. In these interviews, students turned off their videos to keep their identities anonymous. First, students were provided with the unscaffolded pretest followed by the scaffolded pretest. After that, they engaged with the tutorial and then the unscaffolded posttest followed by the scaffolded posttest. The average time for interviews was 3 h 40 min per student for thirteen interviewed students (due to the length of each interview, each student was interviewed in multiple sessions within a given week). We note that this tutorial is intended for advanced undergraduate students but interviews with graduate students also provided valuable insight into student understanding and the effectiveness of the tutorial in helping them learn. We did not find significant differences in the performance of graduate students and undergraduates. These interviews helped us in refining the tutorial along with feedback from four faculty who teach upper-level E&M or other upper-level physics courses.

*3.4. Course Implementation and Instructor Details*

The tutorial was implemented at a large research university in the US. At this university, upper-level E&M is a sequence of two semester courses. The various methods for calculating electrostatic potential including Laplace's equation approach are part of the first course. Although the pretest and posttest were always administered in-class, the tutorial was administered as homework over four consecutive academic years under three different instructors: instructor 1 (I1), instructor 2 (I2), and instructor 3 (I3). In the first three years of implementation, the pretest, tutorial and posttest were administered in the first semester course after traditional lecture-based instruction in relevant concepts. In the fourth year, it was implemented in I3's course at the beginning of the second semester E&M course. I1 taught the upper-level E&M course for two consecutive years and I2 taught it for one year after that. In the fourth year of implementation, I3 taught the second course of the upper-level E&M in the next semester. All students received instructions from the textbook by Griffiths [89]. The number of students present for each test (all students) for each instructor and the number of matched students present (matched students are those who took both pretest and posttest) are shown in Table 2.

**Table 2.** In each instructor's class, the number of students who took the pretest and posttest and the number of matched students (those who took both pretest and posttest) are shown.

| | Number of Students | | |
|---|---|---|---|
| **Instructor** | **Pretest (All)** | **Posttest (All)** | **Pretest and Posttest (Matched)** |
| I1 | 52 | 44 | 44 |
| I2 | 12 | 18 | 9 |
| I3 | 15 | 17 | 15 |

Although the effectiveness of the tutorial is likely to be greater when students work on it in class in small groups with some scaffolding provided by the instructor, LE tutorial was administered as homework in all three instructors' courses since they did not want to use class time to have students engage with it. In classes of I1 and I2, the tutorial was given as homework after traditional lecture-based instruction on relevant concepts. In I3's class, the tutorial was given at the beginning of the course since it was the second course of the upper-level E&M. For I1 and I2, the pretest was given after traditional lecture-based instruction to evaluate how much students knew about Laplace's equation method to solve for the electrostatic potential. After the pretest, the tutorial was given as homework and then the posttest was administered to students to evaluate how much they learned from the tutorial. For I3, the pretest was given at the beginning of the course in the first week of the semester. Then the tutorial was given as homework (although I3 emphasized in their class that this was not homework relevant for the course and they will get a small amount of extra credit for trying) and then the posttest was administered to the students. I1 administered the pretest and posttest as part of graded exam or an in-class quiz, I2 gave these tests for graded extra credit. On the other hand, I3 mentioned that the tutorial and tests were part of a research study and were not relevant for their current course (second course of the upper-level E&M which focuses on electrodynamics instead of electrostatics), and they would get a small amount of extra credit for attempting the pretest, tutorial and posttest which were part of the research study. The pretest and posttest were given in the same order for I1 and I3 but were swapped for I2 (in other words, what was given as the pretest was given as the posttest and vice versa). Based on the preference of instructor 1 (I1), the pretest and posttest were given in the scaffolded form to all students. To maintain consistency, the pretest and posttest were given in the scaffolded form for the other two instructors (I2 and I3) as well.

The scaffolded pretest and posttest each consist of six questions. The first question that asks students to identify whether solving LE by separation of variables is an effective approach for the three situations provided. The other five questions are based on a problem which is broken down into multiple sub-problems to guide students through the problem-solving process (see Appendix A). While the specific problems in the pretest and posttest differ slightly, they are of comparable difficulty. The scaffolded assessments are designed to evaluate students' ability to identify when solving Laplace's equation is an appropriate and effective approach for determining the electrostatic potential, describe the method of separation of variables, express boundary conditions mathematically, determine the correct form of separable solutions, write the general solution as a superposition of separable solutions and find expansion coefficients by carrying out the integration using the Fourier trick to obtain the unique solution.

The tutorial provides scaffolding support to students in developing these skills by guiding them through the process of drawing diagrams of the physical situation, explicitly writing Laplace's equation in the relevant coordinate system, and applying boundary conditions to find separable solutions. It also helps students construct the general solution as a linear combination of separable solutions and compute the expansion coefficients using the Fourier trick to arrive at the unique solution. Although the scaffolded pretest and posttest do not assess many of these skills (since the tests are broken into sub-problems), if students are administered the unscaffolded version (as in all of the interviews), they will be able to demonstrate the full range of abilities (listed in the learning objectives) developed through the tutorial.

To ensure interrater agreement, approximately 24% of the pretests and posttests were graded independently by two graders, and the overall disagreement was found to be between 12% and 0% across different questions. The grading rubric for the latest version of the pretest and posttest are included in the Appendix A.

*3.5. Data Analysis*

After in-class implementation, student performance was analyzed by calculating the average scores along with standard errors on scaffolded pretest and posttest. In addition, for thirteen interviewed students, average scores were calculated separately for the scaffolded and unscaffolded versions of the test. Each question was scored 0–100% based on the rubric provided in the Appendix A. We calculated the effect sizes using Cohen's $d$ [90], which is defined as $d = \frac{posttest - pretest}{\sigma_{pool}}$. Since we used the matched data (students present in pretest as well as posttest) for the detailed analysis, $\sigma_{pool}$ is defined as $\sigma_{pool} = \sqrt{((s_{pretest}^2 + s_{posttest}^2)/2)}$, where $s_{pretest}$ and $s_{postest}$ are the standard deviations of pretest and posttest, respectively. We calculated the effect size to focus on the practical significance, which would be useful for instructors. For the average scores, statistical significance was evaluated using *t*-test and *p*-values are reported.

## 4. Results and Discussion

Although our primary focus was to use student difficulties found via research as a guide to develop and validate the tutorial to help students learn to apply Laplace's equation to solve electrostatic problems, the study revealed additional findings. We observed that interviewed students' learning gains after engaging with the tutorial were more evident on the unscaffolded tests compared to the scaffolded ones, suggesting that the scaffolding provided in the tutorial may have masked the demonstration of different aspects of their learning. Additionally, the framing of instructor 3 (I3), who stated that the tutorial was not relevant for the course syllabus and was part of a research study, may have negatively influenced student engagement and performance in that course (even on the pretest). In particular, although the tutorial was given in a manner that one would traditionally call the first homework at the beginning of the course, the instructor assured students it was not actual homework that they should worry about, and they would obtain a small amount of extra credit for completion regardless of correctness.

Below, we first discuss the difficulties students encountered while working through the tutorial and the corresponding tests on solving Laplace's equation. We describe how specific features of the tutorial were designed to address these challenges and how we iteratively refined the tutorial based on observations of student reasoning and performance to better support their learning.

*4.1. Student Difficulties and How the Tutorial Addresses Them*

4.1.1. Difficulty in Recognizing That Laplace's Equation Applies for a Given Problem

During interviews, we found that a majority of students struggled to recognize that the problem presented in the unscaffolded pretest could be effectively solved using Laplace's equation (LE). Although the title of the pretest mentioned LE, many students did not notice this and did not use LE for solving the problem. For example, among the thirteen students interviewed, two incorrectly thought that the method of images would be the appropriate approach for determining the potential, even though this method is not applicable for the given configuration. In contrast, students' posttest performance in the unscaffolded version (as shown in Figure 1) shows clear improvement. Figure 1 shows the average pretest (PreU, solid fill) and posttest (PostU, hatched filled) along with standard errors and effect size (Cohen's $d$ [91]) for the interviewed students. After engaging with the tutorial, students were able to correctly identify the application of LE and approached the problem with greater confidence and accuracy.

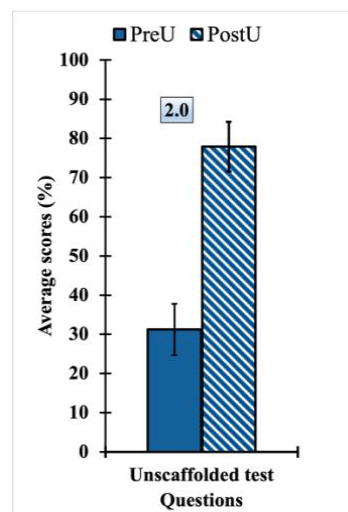

**Figure 1.** Average percentage scores of 13 interviewed students on the unscaffolded pretest (PreU) and posttest (PostU). The standard errors and effect size (Cohen's $d$) are also shown.

During the interviews, in the scaffolded tests, students appeared to recognize the use of LE more readily, likely because it was explicitly stated in the problem. Q1 of the scaffolded tests and the tutorial presented three different configurations, only one of which could be appropriately solved using LE. Students were asked to identify which configuration(s) would be suitable for solving LE effectively, and they were allowed to select more than one configuration given. The performance on this task during interviews revealed that many students struggled to identify the correct configuration(s) in which LE is effective. Some relied on incorrect reasoning. For example, one student concluded that LE should be used whenever electric field lines appeared to curve or curl and said, '*I just thought…that it was any time there's some 2d curvature in the E lines* [field lines curve or bend] *is when we're supposed to use Laplace's equation*'. This student drew electric field lines for each configuration to incorrectly conclude whether LE can be effectively used. Several interviewed students thought that LE could be used in multiple scenarios given, including those involving charges in the region of interest, which is incorrect. Some suggested that perhaps LE can also be used for some configurations in which they are usually taught to use the method of images.

Based on the difficulties observed in the interviews, we revised the tutorial. In addition to the tutorial asking students to select the situations in which LE is appropriate and effective for finding electrostatic potential, we introduced a dialogue between two hypothetical students discussing common incorrect choices and reasonings. For example, in the following question we introduced a dialogue between two students that highlights a common incorrect reasoning students provide while selecting the situations in which LE is applicable. One student (Sarah) selects an incorrect situation for the problem provided in the tutorial and offers reasoning that is frequently observed in student interviews. The other student (John) disagrees and provides a correct explanation. Students engaging with the tutorial are asked to contemplate these and decide with whom they agree as follows:

> **Sarah:** *We can use Laplace's equation for all three situations mentioned in Q1 (problem given in the tutorial). Although in option I (a point charge present near a grounded conducting plane), the method of images could be an effective method, solving Laplace's equation to find the potential could also be an effective approach.*
>
> **John:** *I disagree. Laplace's equation $\nabla^2 V = 0$ is only applicable in situations where there is no charge density present within the region. Since for situations I and III (a thin disc with a uniform charge distributed on it) of Q1, there is non-zero charge density present in the region where we want to solve for the potential, we cannot use Laplace's equation.*

This dialogue highlights the necessary and sufficient conditions for applying LE: specifically, that the region of interest must be charge free and that all relevant boundary conditions must be specified. Later, after the students had the opportunity to think about these issues, they also encountered checkpoints to support reflection and reinforce conceptual understanding such as the following:

> **Checkpoints:**
> - *Laplace's equation can be used to find the potential for a well-defined boundary value problem having no charge density in the region of interest (enclosed by the boundary).*
> - *The boundary conditions must be known to obtain the unique solution for the potential in the specified region.*

These refinements appeared to be effective; in the later interviews, we observed improvement. For example, one student initially identified a scenario in the scaffolded pretest Q1, incorrectly claiming that Laplace's equation would be appropriate and effective even when it was not. However, after engaging with the tutorial, the student was able to correctly identify the applicable scenario in the scaffolded posttest based on the appropriate criteria for using LE.

4.1.2. Difficulty in Using the Method of Separation of Variables and Finding Separation Constants

During the interviews, in the unscaffolded pretest, we observed that most students who were able to figure out that Laplace's equation was applicable to solve the given problem did not make further progress by explicitly applying the method of separation of variables to find the separable solution. Furthermore, in the scaffolded tests, in Q2, students were asked to explain the method of separation of variables. The rubric for this question was designed to assess students' understanding of the method of separation of variables. Students would earn full points by providing a complete description of the process, including the application of the Fourier trick to determine expansion coefficients and describing how to find the unique solution from the general solution. However, in the scaffolded pretest, many students could not provide a response that was sufficient to receive full credit based on rubric. This is evident, e.g., from the case of a student who applied the method of separation of variables to solve Laplace's equation and scored 100% on the unscaffolded posttest. However, when presented with the scaffolded posttest immediately after that, this interviewed student scored only 67% for Q2. The response of the student for Q2 on the scaffolded test was, '*A PDE [partial differential equation] is difficult to solve but if we assume $F(x, y, z)$ is separable, i.e., $F(x, y, z) = X(x)Y(y)Z(z)$, and we get it to be separable, i.e., no mixed derivatives like $\frac{\partial X}{\partial y}$, then it simplifies to ODE [ordinary differential equations] that are set equal to constants because the functions must vary independently yet still be equal (which means they must be constant).*' The students did not achieve full credit on this question because discussion of other essential components, Fourier trick and determination of coefficients were missing according to the rubric. This instance highlights that even when the student has gained the ability to apply the method, the scoring on Q2 due to the strict rubric led to lower score.

In Q4 of the scaffolded pretest and posttest, students were given the option to choose the separable solution they would obtain after applying the separation of variables and appropriate boundary conditions. However, we note that in these scaffolded pre and posttests' Q4, the correct solution can be identified by skipping the method of separation of variables and simply checking if the boundary conditions are being satisfied in each separable solution given in the choices. Therefore, we could not figure out how much students learned about the separation of variables by grading the scaffolded posttest. But the facility of students in applying the separation of variables correctly so that they can proceed further to find the potential was observed in the interviews from the performance of students on the unscaffolded version as shown in Figure 1. Although most students did not talk in detail about the solution process in the unscaffolded or scaffolded pretest during interviews, the unscaffolded posttest shows that many students could successfully implement the appropriate approach to solving the problem, even if they did not articulate them earlier in the pretest. The scaffolding provided in the tutorial supported students in choosing the relevant coordinates and writing Laplace's equation correctly. After students separate the differential equation with the $z$ variable from the one with the $x$ and $y$ variables, they are asked to evaluate the validity of a hypothetical student's (Emily) statement, by agreeing or disagreeing and providing the reasoning, to help further clarify how a partial differential equation can be transformed into ordinary differential equations and how separation constants are defined as follows:

**Emily:** *The equation $\frac{1}{X}\frac{d^2X}{dx^2} + \frac{1}{Y}\frac{d^2Y}{dy^2} = c$ can be broken down further into $\frac{1}{Y}\frac{d^2Y}{dy^2} = c_1$ and $\frac{1}{X}\frac{d^2X}{dx^2} = c_2$ (where $c_1, c_2,$ and $c$ are constants and $c_1 + c_2 = c$) using an* *argument similar to the one we used to separate the z variable. This way we would obtain three ordinary differential equations or ODEs for each variable x, y, and z.*

In this case, Emily is correct and the scaffolding provided with this type of reflective task can help students with these issues.

In part II of the tutorial, a three-dimensional problem prompts students to reflect on and apply what they have learned in earlier parts of the tutorial. They are initially asked to solve the problem on their own as follows:

*Use separation of variables in the Cartesian coordinates to find the solutions of Laplace's equation in the form $X(x)Y(y)Z(z)$. Obtain three different ODEs, one each for $X(x), Y(y)$ and $Z(z)$. Explain any pertinent steps. (Note that the separable solutions in this problem are different from the previous problem you solved).*

After students attempt to solve the problem, the scaffolding support is provided in the next part to help them reflect on and reconcile any possible discrepancies between their initial solution and the guided solution. We observed during interviews that students appeared to be more confident while using LE in the three-dimensional problem which also provided an opportunity to students to reflect upon their understanding of the method of separation of variables. Overall, both the tutorial and the unscaffolded posttest responses suggest that most students developed a good understanding of the necessary conditions for applying Laplace's equation to appropriate problems and also learned to apply the method of separation of variables in the context of finding the electrostatic potential.

4.1.3. Difficulties with Boundary Conditions

Solving Laplace's equation (LE) requires students, for example, to use the boundary conditions to determine appropriate separable solutions. Although the scaffolded tests did not explicitly ask students to solve for separable solutions, Q4 in the scaffolded pre and posttests assessed this issue indirectly by asking students to select the correct functional form from multiple-choice options. We find that most students made their choices based on how the proposed functions satisfied the boundary conditions. Many students struggled in finding the correct separable solution in the Q4 of pretest during interviews. As a result, several problems in the tutorial were intentionally designed to highlight the importance of boundary conditions in selecting the form of the solution and determining the correct form of the differential equation. Students were also asked to apply the boundary conditions to find different constants. In early interviews, students frequently struggled to apply the correct boundary conditions when solving for the coefficients in the separable solutions. The tutorial is designed to help students develop a coherent understanding of the boundary conditions for the given configuration so that they can translate, e.g., between the verbal and mathematical representations. For example, students were asked to write the boundary conditions mathematically for the given configuration as follows:

*Let's write down the boundary conditions for this problem mathematically.*

*(a) What is the potential at $x = 0$?*
*(b) What happens to the potential as $x \to \infty$?*
*(c) What is the potential at $y = 0$?*
*(d) What is the potential at $y = a$?*
*(e) At fixed values of the x and y coordinates within the channel, does the potential depend on the z coordinate?*

Students are also scaffolded by asking them to consider the behavior of the function based on boundary conditions for the given configuration as follows:

*Answer the following questions:*

*(a)* Is $X(x)$ oscillatory as a function of $x$ for the given boundary conditions?

*(b)* Is $Y(y)$ oscillatory as a function of $y$ for the given boundary conditions?

These questions provide scaffolding support to help students learn to write the boundary conditions mathematically. The learning sequences in the tutorial guide them to analyze how the different functions used in solving the problem behave based on the boundary conditions.

The tutorial also incorporates conversations between students that highlight common difficulties in interpreting and applying the boundary conditions. For example, in the following hypothetical discussion between students while writing ordinary differential equations, a student, John, appears to be confused about why a specific sign convention is chosen for the given boundary value problem when his friends discuss using the separation of variables to turn a partial differential equation into ordinary differential equations. In the tutorial excerpt discussed below, the conversation continues based on a disagreement between students in which Emily and Sarah try to help resolve John's confusion. Students engaging with the tutorial are expected to agree or disagree with these students and provide their reasoning in each case:

> **John:** *I still don't understand why a positive sign in front of the constant $k^2$ for the differential equation involving the function $Y(y), i.e., \frac{d^2Y}{dy^2} = k^2Y$, and a negative sign in front of the constant $k^2$ for the differential equation involving the function $X(x), i.e., \frac{d^2X}{dx^2} = -k^2X$, would not work.*
>
> **Emily:** *In this problem, consistent with the boundary condition, $X(x)$ involving the x coordinate must decay to zero far away from the $x = 0$ plane, i.e., as $x \to \infty$. That means the solution $X(x)$ involving the $x$ coordinate must involve the decaying exponential function. This means we must choose a positive sign in front of the constant $k^2$ for the differential equation involving the function $X(x)$.*
>
> **Sarah:** *I agree with Emily. Boundary conditions require $Y(y)$ to have at least two zeros, and only the oscillatory functions can satisfy these boundary conditions out of the given choices of solution. Therefore, we should choose a negative sign in front of the constant $k^2$ for the differential equation involving the function $Y(y)$.*

During the interview, many students had difficulty in finding the constants, but as they proceeded through the tutorial, they became comfortable with this task. Part II of the tutorial included a three-dimensional problem which is comparatively challenging but in which less scaffolding support is provided to help students develop self-reliance. After engaging with the tutorial, students showed a better understanding of how to choose appropriate functional forms based on given boundary conditions in the scaffolded posttest. Based on students' struggles, we also introduced checkpoints and made targeted modifications to the tutorial which helped in improving students' conceptual understanding. Overall, our findings suggest that the tutorial is effective in deepening students' understanding of the boundary conditions and their role in solving problems involving Laplace's equation.

4.1.4. Difficulty in Finding General and Unique Solutions

During the interviews, in the unscaffolded pretest, only two out of thirteen students were able to obtain the separable solution for the given problem correctly. However, none of them appeared to recognize that the general solution should be expressed as a sum of separable solutions. In the scaffolded pretest Q5, a hint was provided that the general solution should be written as a series expansion. However, many interviewed students demonstrated uncertainty about how and why to use a series in this context. During interviews, when students were prompted to find the coefficients $C_n$ in the tutorial,

14 of 34

several were confused, and asked why such coefficients were necessary at all. One student said, '*I don't understand why we have a $C_n$...we're summing $C_n$ from n equals one to infinity... I don't know why we would sum in a potential*'. Although some of the students identified $k_n$, they were still confused about why there should be summation over all values of $n$ while writing the general solution. When students were asked to choose the correct form of the general solution, in the following question, some selected two options, unsure if a sum was required, indicating uncertainty about whether a linear superposition of separable solutions was appropriate:

> *Circle all the correct answers regarding the form of the general solution for Laplace's equation obtained using linear superposition of separable solutions (which were found by imposing some of the boundary conditions to obtain the separation constants):*
>
> I. $V = \sum_{n=1}^{\infty} C_n e^{-n\pi x/a} \sin\left(\frac{n\pi y}{a}\right)$
> II. $V = C e^{-n\pi x/a} \sin\left(\frac{n\pi y}{a}\right)$
> III. $V = \sum_{n=1}^{\infty} \sum_{\substack{m=1 \\ m \neq n}}^{\infty} C_{mn} e^{-n\pi x/a} \sin\left(\frac{m\pi y}{a}\right)$

Despite having learned about partial differential equations in different physics courses as well as in the upper-level E&M course, many students appeared to be unclear about why the superposition via the summation is relevant in this physics context. To address this conceptual gap, in the tutorial, we added a conversation between three students (John, Emily, and Sarah) as follows, discussing the role of superposition and the reasoning for expressing the solution as a sum of the separable solutions which appeared to be effective in later interviews:

> **John:** *$V = Ce^{-kx}\sin(ky)$ is the unique solution for the potential in the channel for the given problem, where $k = \frac{n\pi}{a}$.*
>
> **Emily:** *I disagree. Since $k$ is dependent on n, $C_1 e^{-\frac{\pi x}{a}}\sin\left(\frac{\pi y}{a}\right)$, $C_2 e^{-\frac{2\pi x}{a}}\sin\left(\frac{2\pi y}{a}\right)$, or $C_6 e^{-\frac{6\pi x}{a}}\sin\left(\frac{6\pi y}{a}\right)$, etc., can all be solutions of Laplace's equation if $k$ takes the value $\frac{\pi}{a}, \frac{2\pi}{a}$ or $\frac{6\pi}{a}$, respectively. A linear combination of these solutions with n = 1, 2...∞ is a general solution for Laplace's equation with the boundary conditions used so far to obtain the separable solutions.*
>
> **Sarah:** *I agree with Emily. The integer $n$ in $k_n = \frac{n\pi}{a}$ can take any integer value from 1 to ∞. Therefore, each of these separable solutions obtained can be written as $V_n = C_n e^{-k_n x}\sin(k_n y)$. The general solution can be written as a linear combination of these separable solutions, $\sum_{n=1}^{\infty} V_n$. "General solution" means that this series solution applies to a family of boundary value problems which all satisfy the boundary conditions already applied but with one boundary condition yet to be applied.*
>
> **Emily:** *I agree with Sarah! We can then obtain the unique solution of Laplace's equation, which is the potential in the channel, when we find the coefficients in the linear superposition of separable solutions using the Fourier trick exploiting the remaining boundary condition.*

In the problem discussed here, John expresses the confusion related to the form of general solution which eventually gives the unique solution which was commonly seen during interviews. Emily and Sarah provide reasoning about how the general solution should be written. Interviews suggest that the conversation among students and the additional checkpoint helped students evaluate and reflect upon their reasoning. We observed that students were more comfortable with the reasoning behind the need for a summation in the general solution and answered confidently in the later part of the tutorial when they had to choose the correct form of the general solution. In the unscaffolded posttest, most of the students were able to correctly write the general solution as a sum of separable solutions, suggesting a significant improvement in their conceptual understanding. This implies that the difficulty related to writing the general solution decreased as students engaged with the tutorial.

4.1.5. Difficulty in Finding the Expansion Coefficients $C_n$

Once the general solution is obtained, students must find the unique solution to determine the potential in the region of interest. Therefore, it is important to find the expansion coefficients $C_n$ to arrive at the unique solution that satisfies the remaining boundary conditions. During the interviews, in the unscaffolded pretest, none of the students applied the Fourier trick to solve for the constants to find the unique solution. In the scaffolded test, students were given the integral form of the equation they were asked to solve, which was obtained by applying the Fourier trick in scaffolded pretest Q6. Thus, the scaffolded test did not ask students to explicitly apply the Fourier trick to find the solution. Therefore, we could not observe the extent to which students learned to apply the Fourier trick by engaging with the tutorial. Via hypothetical students' conversations, e.g., between John and Emily as illustrated below, in which students are asked to agree or disagree with each person and provide reasoning, the tutorial helps students learn how to apply the Fourier trick to find the expansion coefficients, which are essential for determining the unique solution:

*John: We are completely done finding the unique solution for Laplace's equation with the potential* $V(x,y) = \sum_{n=1}^{\infty} C_n e^{-n\pi x/a} \sin\left(\frac{n\pi y}{a}\right)$.

*Emily: I disagree. We still must find the coefficients* $C_n$. *We should use our remaining boundary condition (5a)* $V(0,y) = V_0 = constant$ *at* $x = 0$ *and Fourier's trick to solve for the coefficients* $C_n$.

In this problem, Emily correctly hints that the unique solution is still not determined unless the $C_n$ are found using the Fourier trick. We observed an improved proficiency in students' ability to apply the trick and find the potential as they proceeded to part II of the tutorial. The facility with which students were able to find the coefficients correctly in the unscaffolded posttest during the interviews shows that most of the interviewed students were able to apply the Fourier trick confidently. Not only that, but students also realized that finding the expansion coefficients helps in obtaining the unique solution which represents the electrostatic potential in the region of interest. Overall, the tutorial helped reduce student difficulties, and iterative changes further refined and enhanced its effectiveness.

4.1.6. Student Engagement with Visual Representations Aids in Problem-Solving

In addition to student difficulties discussed above, interviewed students showed varied levels of engagement with visual representations [92], which plays a key role in reasoning about the problem. Therefore, the tutorial included explicit prompts asking students to draw diagrams. Interviews suggest that these prompts, as in the following example, supported their understanding and helped reduce difficulties:

*"Draw a figure of this situation using the coordinate system you chose in the previous problem. Place the semi-infinite grounded conducting sides at* $y = 0$ *and* $y = a$. *Indicate the known potentials on the three sides of the channel (draw a 2D figure instead of a 3D figure)."*

Additionally, the tutorial included a challenging task in which students are asked to draw the electric field lines and equipotential surfaces as in this example:

*"Make a sketch of equipotential surfaces and electric field lines within the slot. You can do this using conceptual reasoning. Is the analytic solution for the potential expressed as an infinite series helpful in making the sketch?*

*Hint: Electric field lines intersect at the boundaries of a conductor perpendicularly. Equipotential surfaces are perpendicular to the electric field lines."*

Students were explicitly asked to draw diagrams three times in the tutorial, but many chose to draw additional diagrams on their own. The structure of the tutorial, with adequate space and opportunities for reflection, appears to encourage students to draw diagrams, especially when reasoning about the boundary conditions or analyzing behavior of the different functions representing mathematical solutions of the differential equations, electric potential or electric field lines.

Building on these insights, primarily from interviews, we now discuss quantitative findings from classroom implementation to evaluate the tutorial's effectiveness and reflect upon challenges encountered in the in-class implementation.

*4.2. Results from Classroom Implementation*

We now compare the pretest and posttest performance of students across three instructors for the scaffolded pretest and posttest (no unscaffolded tests were used for in-class implementation). Since there was a small number of students for instructor 2 (I2), we wanted to make sure that the analyses were consistent for all students (those who took only the pretest, posttest, or both) and only matched students (those who completed both pretest and posttest). The numbers of all students and matched students for I2 are shown in Table 2. Figure 2 shows the average percentage scores on pretest (hatched fill) and posttest (solid fill) along with error bars (standard error) for students of I2, comparing all students (I2PreAll and I2PostALL, left pair) and matched students (I2PreM and I2PostM, right pair) across all six questions provided in the scaffolded tests. Numbers in boxes above the pair of each pretest and posttest indicate the effect sizes (Cohen's *d*) for all and matched students across all six questions. Figure 2 reveals that the average scores for all students and matched students are remarkably similar, confirming the robustness of our sample. Notably, no significant differences emerged between all vs. matched groups for I2, a pattern consistently seen with I1 and I3 as well. This consistency motivated us to conduct the rest of the analysis and comparison with matched students for all three instructors. From this point on, we refer to them simply as students.

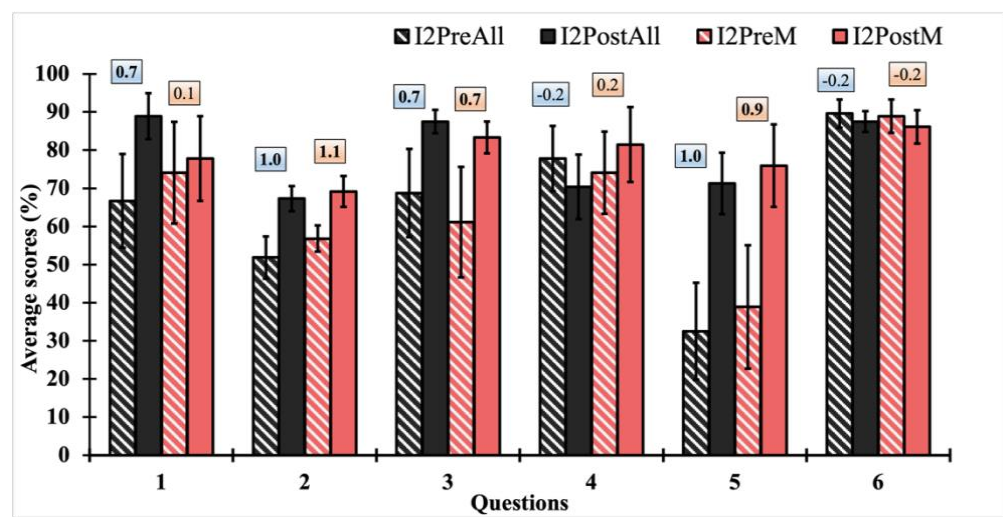

**Figure 2.** Average percentage scores on Laplace's equation pretest and posttest for instructor 2 or I2 for all (I2PreAll and I2PostAll, left pair) and matched (I2PreM and I2PostM, right pair) students along with standard errors and effect sizes (Cohen's *d*) from pretest to posttest.

Figure 3 displays the average percentage scores with error bars on the pretest and posttest for instructors I1 (left pair) and I2 (right pair) across all six questions from the scaffolded tests. For I1, the pretest is shown with a solid fill and the posttest with a hatched fill for each question. For I2, pretest and posttest were swapped compared to I1 and therefore in the right pair, the pretest is represented by hatched fill and the posttest is represented by solid fill for each question. Similarly, Figure 4 shows the average % scores with error bars on the pretest and posttest for instructors I1 (left pair) and I3 (right pair) across all six questions provided in the scaffolded tests. The order of tests was the same for I1 and I3 and therefore, in both the left (for I1) and right pairs (for I3), the pretest is represented by solid fill and the posttest is represented by hatched fill for each question.

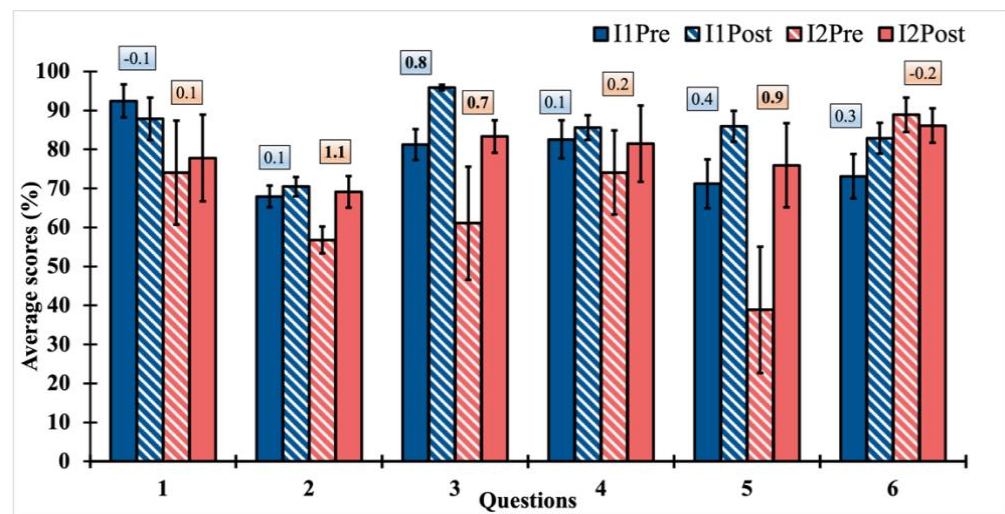

**Figure 3.** Average percentage scores of students on scaffolded pretest and posttest on Laplace's equation for instructor 1 or I1 (I1Pre and I1Post, left pair) and instructor 2 or I2 (I2Pre and I2Post, right pair) along with standard errors and effect sizes (Cohen's *d*) from pretest to posttest.

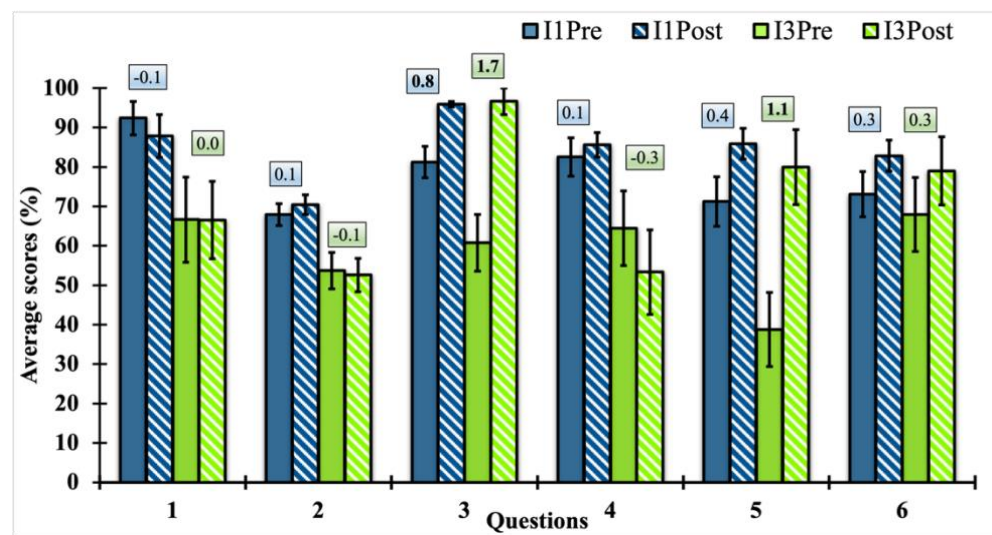

**Figure 4.** Average percentage scores of students on scaffolded pretest and posttest on Laplace's equation for instructor 1 or I1 (I1Pre and I1Post, left pair) and instructor 3 or I3 (I3Pre and I3Post, right pair) along with standard errors and effect sizes (Cohen's *d*) from pretest to posttest.

In the scaffolded version, Q1, Q2 and Q4 in the pretest and posttest evaluate students' ability to identify the situations in which Laplace's equation applies, describe the method of separation of variables, and find the type of functions that should be chosen for the given boundary condition. Q3, Q5 and Q6 focus on students' ability to write the boundary conditions mathematically, write the general solution and carry out the integration involving the Fourier trick to find the expansion coefficients.

Converting the verbal representation to mathematical representation is a challenging task, and we observed that many students initially struggled with this, particularly when writing the boundary conditions for a given problem. Figure 3 and Figure *4* show notable improvements in student performance on Q3 across all three instructors, suggesting that the tutorial supported students in developing better understanding of how to represent the boundary conditions mathematically.

Another noteworthy improvement evident in student performance on the scaffolded test for all three instructors is the ability to express the general solution to Laplace's equation as an infinite sum of separable solutions. This skill is assessed in Q5 and reflects a key part in applying the method of separation of variables. The gains observed across all instructors suggest that the tutorial effectively supported students in understanding the structure of the general solution.

While Figure 3 and Figure *4* show clear improvements in areas such as representing the boundary conditions mathematically and constructing the general solution, the performance on other questions show limited gains.

Q1 assesses whether students can identify the situations in which using Laplace's equation is an effective approach for finding electrostatic potential. For I1, students were already performing well on this question in the pretest (see Figure 3) and this performance was sustained in the posttest. For I2, we observe gains from the pretest to posttest (see Figure 3), but they are not significant. In contrast, students in I3's class (see Figure 4) did not show any change for this question.

In Q2 of the scaffolded tests, students were also asked to explain the method of separation of variables. We expected students to not only mention turning a partial differential equation into ordinary differential equations but also describe key components such as the use of the Fourier trick, the form of the general solution and constraints on constants. However, in retrospect, we think that our grading rubric for this question may have been overly strict. Many interviewed students appeared to understand these ideas well enough to implement them in problem-solving, even if they did not articulate them fully in writing. As a result, their conceptual understanding may not be accurately reflected in their written responses and associated scores for Q2.

We do not observe significant improvements in Q4 and Q6 in the scaffolded tests for all three instructors as shown in Figure 3 and Figure *4*. Although these questions target key aspects involving solving LE, the scaffolded test did not require students to explicitly demonstrate these proficiencies. In Q4, students were asked to choose the correct form of the separable solution for the given boundary conditions, but they were not explicitly required to solve the differential equation to find the separable solutions. Students focused on verifying the boundary conditions to identify the relevant function, but this approach limited their opportunity to engage deeply with the underlying process such as transforming the partial differential equation into ordinary differential equations and then finding separable solutions.

Similarly, Q6 in the scaffolded tests only evaluates students' ability to compute the integral derived from applying the Fourier trick. Students of I2 appeared to be doing well on the problem already (see Figure 3) and they maintained their performance in the posttest. There has been some improvement in student performance for I1 and I3 (see Figure 4) but it is not significant. However, as noted, in this scaffolded version of the tests, the performance on Q6 does not capture their conceptual understanding of how the general solution can be used to find the unique solution with the help of the Fourier trick and the determination of the expansion coefficients.

Overall, the scaffolded test results suggest that students made noticeable gains in specific skills, such as representing the boundary conditions mathematically (Q3) and expressing the general solution as a sum of separable solutions (Q5). However, improvements were limited in other critical areas, including recognizing when Laplace's equation is applicable (Q1), explaining the method of separation of variables (Q2), selecting the separable solution (Q4), and performing integration. This does not necessarily mean that students did not learn from the tutorial; rather, it suggests that the scaffolded assessments may not have fully captured the extent of their conceptual understanding in these areas as discussed earlier (see Figure 1 for the pre-/posttest differences for the interviewed students on the unscaffolded version of the tests).

To better understand the extent of student learning and how scaffolding may have influenced their reasoning, we look at the performance of the interviewed students, who completed both the scaffolded and unscaffolded (Figure 1) versions of the test.

Figure 5 shows that the overall performance of the interviewed students on each question of the scaffolded pretest and posttest closely mirrors the in-class performance of students for all three instructors. However, these same interviewed students demonstrated substantial gains on the unscaffolded test as shown in Figure 1, which they completed prior to the scaffolded versions of the corresponding test. This pattern suggests that the interviewed students had meaningfully learned from the tutorial as evidenced by their ability to solve the problem independently on the unscaffolded posttest, yet the gain was not captured by the scaffolded posttest. The breaking of the problem into sub-problems in the scaffolded pretest and posttest masked evidence of learning. In particular, in the scaffolded versions, the problems were already broken into sub-problems, so students had only limited opportunity to demonstrate what they had learned.

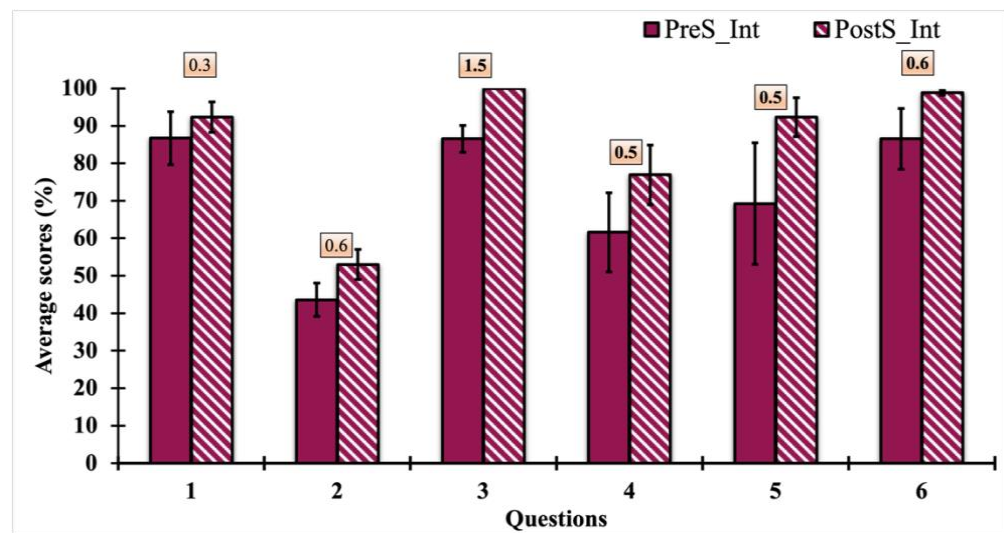

**Figure 5.** Average percentage scores of interviewed students on scaffolded pretest (PreS_Int) and posttest (PostS_Int) on Laplace's equation along with standard error and effect sizes (Cohen's *d*) from pretest to posttest.

The average scores of students on Q2–Q6 of the scaffolded test, across all three instructors (I1S, I2S, and I3S) and the interviewed students (IntS), and their comparison with the average scores of interviewed students on the unscaffolded test (IntU) are shown in Figure 6. We conducted *t*-test and found that the corresponding $p$ values for the students of all three instructors and interviewed students are I1S ($p < 0.001$), I2S ($p =$ 0.036, I3S ($p = 0.013$), IntS ($p = 0.059$), and IntU ($p < 0.001$). Thus, these performance differences between the pretest and posttest are either statistically significantly different or approaching statistical significance ($p < 0.05$) despite the number of students being small. Q1 is not included in calculating the average as it is not a part of the problem given in unscaffolded form. This comparison shows that the student performance was comparable across all groups for the posttest (except I3's class having somewhat lower posttest performance). Figure 6 shows that the average scores on the scaffolded pretest, in general, were much higher than that on the unscaffolded pretest (interviewed students only). This shows that student performance on the scaffolded pretest was likely boosted due to advantages received from having the problem broken down into sub-problems. The scaffolded test provided support to students who may have otherwise not have been able to solve the problem independently, especially in the pretest after traditional instruction without the problem being broken down into sub-problems. Thus, the scaffolded pretest masked our ability to accurately evaluate students' actual understanding of the underlying concepts and problem-solving.

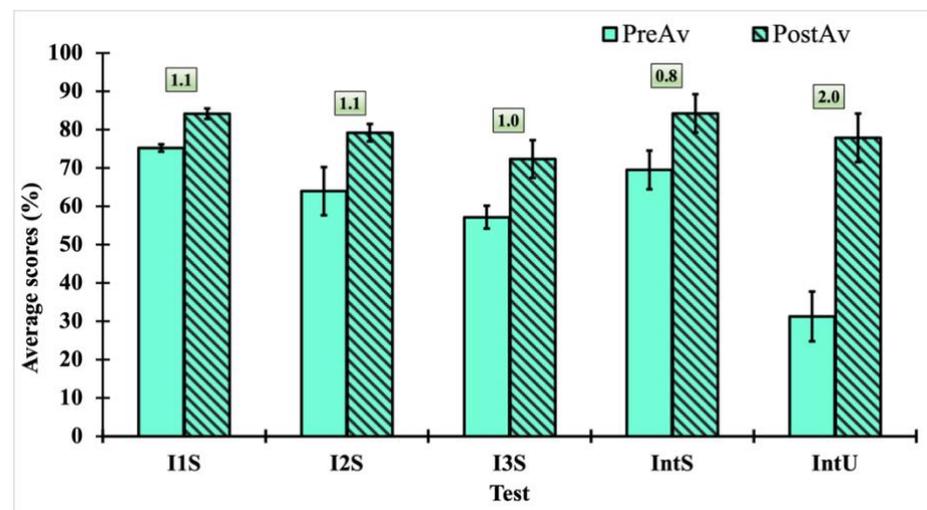

**Figure 6.** Average percentage scores (Q2–Q6) on scaffolded pretest and posttest for instructor 1 or I1, instructor 2 or I2, instructor 3 or I3, and interviewed students and on unscaffolded pretest and posttest for interviewed students along with standard error and effect sizes (Cohen's *d*).

*4.3. Reflection on the Scaffolding Paradox in Assessment*

These findings highlight a key drawback of the scaffolded assessment format. Although scaffolding can support students in problem-solving, it may also obscure the depth of their conceptual understanding. In this study, the unscaffolded test employed in interviews (in addition to scaffolded tests) revealed that students had indeed learned from the tutorial demonstrating the ability to independently apply Laplace's equation whereas the scaffolded version of the tests did not fully capture these gains. In fact, the structure of the scaffolded tests may have inflated the performance of students who could not solve the problem independently, particularly on the pretest, by offering sub-problems that allowed students to progress without showing the full grasp of the underlying concepts.

*4.4. Reflection on the Impact of Instructor Framing on Student Engagement*

Apart from these overall patterns, we also observed that students in I3's class did not show meaningful improvement from the pretest to posttest. Importantly, their performance on the pretest itself was lower than other groups' pretest performance. We note that I3 emphasized that these tests and tutorial were a part of a research study in physics education and were otherwise not relevant for the course, and students were given a small amount of extra credit for attempting them but not for correctness. This type of framing from the instructor and lack of incentive to learn could have impacted student engagement and performance in I3's class. In particular, since I3 mentioned to students that these tests and tutorials were part of a research study and were not relevant to the current syllabus of the course, it may have reduced the incentive to actively learn from the tutorial and participate in the assessment tasks. In particular, the lower performance observed in I3's class could potentially signify that students did not meaningfully engage with the pretest, tutorial and posttest to learn these concepts and did not gain on related learning objectives.

## 5. Broader Discussion

This study contributes to ongoing efforts in physics education research to understand and support student learning in upper-level E&M courses. We sought to identify common student difficulties and used them as a guide for the development, validation, and implementation of the research-based tutorial to solve for the electrostatic potential using LE. Additionally, we explored the impact of breaking the pre-/posttest problems into sub-problems (unscaffolded vs. scaffolded versions of the tests) and instructor framing of instructional tasks on student performance.

Based upon performance gains from the pretest to posttest, we find that overall, students became more facile at different aspects of solving physics problems involving Laplace's equation, particularly those involving mathematical representation, for all three instructors after engaging with the tutorial. The tutorial appeared to support a consistent use of diagrammatic representations. Even when not prompted, many students chose to draw diagrams, suggesting that they recognized the value of such representations when reasoning about boundary conditions, the behavior of different functions encountered throughout the problem-solving process, and the behavior of electric field lines and electric potential. This practice was supported by the tutorial's structure, which included explicit prompts to draw figures in selected questions and ample opportunity for reflection. Encouraging diagrammatic representation as a part of the problem-solving strategy and providing students with incentives may help both in improving their conceptual understanding and problem-solving abilities.

Another major finding of this research relates to the scaffolding paradox in assessment. Due to the instructor preference and course time constraints, students were administered the scaffolded version of the pretest and posttest. These constraints revealed an important finding that scaffolding in the pre-/posttest, i.e., breaking the problems into sub-problems, may mask evidence of students' independent problem-solving abilities. For example, based upon interview findings, administration of unscaffolded tests in classes would have likely shown better gains from the pretest to posttest in the in-class administration. These findings provide an important insight into how breaking a problem into parts (as in the scaffolded version of the assessment) can artificially inflate performance when students know less (e.g., in the pretest) and mask evidence of growth (e.g., in the posttest). In particular, the extra support provided to students in the scaffolded version of the pretest prevented fully identifying what students did not know or where they lacked understanding (e.g., they may not have been able to break the problem into sub-problems on their own without support). Furthermore, in the posttest, the scaffolded test masked what students might have genuinely learned or gained from the tutorial beyond their knowledge before engaging with the tutorial because they were not required to demonstrate their knowledge in an unscaffolded form (i.e., when scaffolding support has been faded). Thus, the differences observed in the unscaffolded and scaffolded tests of interviewed students reveal the underlying patterns (e.g., significant gain from pretest to posttest after engaging with the tutorial on the unscaffolded version of the assessment) that were obscured by only the scaffolded versions being administered in the class. In particular, "fading" of the scaffold as in the unscaffolded assessment is necessary to obtain a truer baseline of students' independent ability to solve problems without the scaffolding. Consistent with our framework, the goal of instruction is to remain in students' ZPD, which is dynamic and evolves as students learn, and to ensure that students gradually reach a place where the scaffold is no longer needed because they are self-reliant in problem-solving. Our study powerfully demonstrates that a scaffolded assessment with permanent "unfaded" scaffolds fails to accurately measure this growth in students due to their engagement with the tutorial.

Finally, the study also revealed how instructors frame an instructional activity such as a tutorial (and the corresponding tests) and the incentives they provide may influence student engagement and performance. When students view a task as integral to the course and are given meaningful incentives, they are more likely to engage deeply. In contrast, framing it as a research study not relevant for the course and offering minimal extra credit for completeness only may reduce student motivation. The difference among instructors in this study highlights the need for thoughtful incentivization to foster meaningful student engagement in the process of learning.

## 6. Instructional Implications

We note that the tutorial is intended to serve as Supplementary Material rather than as the primary tool for introducing concepts to students. The tutorial uses a guided inquiry-based approach and is carefully designed to develop students' conceptual understanding and problem-solving skills. In this study, the tutorial was used as homework. However, it can be used as an in-class learning tool and students can be asked to work on it in small groups supervised by the instructor. Furthermore, instructors do not even need to implement the full tutorial, they can implement only part of the tutorial by selecting certain parts that are most challenging for their students based on their preference and time-constraints. This flexibility can enable instructors to use the tutorial's guided inquiry-based structure while using only part of the specific problems, provided the part of the tutorial they use with their students remains consistent with the learning objectives. Furthermore, after lecture-based instruction on the concepts, questions from the tutorial can also be adapted into think-pair-share or clicker questions [93-95]. These two approaches are similar, except that think-pair-share does not require feedback through clickers, and its questions may or may not be framed in the form of multiple-choice. Even in the think-pair-share format, however, students can be asked to share what they discussed with their peers, followed by a full class discussion. Further, the tutorial involves multiple conversations between hypothetical students in which common difficulties are raised by one or more students and subsequently addressed by others who provide the correct explanation. These conversations can also be incorporated into class activities, such as think-pair-share or clicker questions, by asking students about which hypothetical student(s) they agree/disagree with. This can be followed by class discussion focused on the underlying concepts.

## 7. Limitations and Future Work

There are several limitations to consider when interpreting these results. The small sample sizes reduce statistical power and limit generalizability. Moreover, all data were collected at a single university, which may not represent the broader populations of physics students. Differences among instructors, each using their own traditional lecture approaches may have influenced students' initial preparation levels. Similar studies conducted at other institutions of different types within a country as well as across different countries around the world would be very valuable.

Furthermore, students engaged with the tutorial as homework, which may limit their level of interaction and collaboration with their peers [95-99]. Embedding the tutorial as an in-class activity, with students working in small groups, could foster deeper engagement and enhance learning outcomes. Future implementations could aim to explore how student performance might change if the tutorial was implemented as a collaborative in-class activity. When students work in small groups during class, the instructor or teaching assistant can move around to ensure that students are engaging productively with one another and making sense of the guided inquiry-based tutorial. It would be instructive to evaluate if such an approach would promote both greater engagement and learning than homework implementation of the tutorial.

**Supplementary Materials:** The supporting information can be downloaded at: https://www.mdpi.com/article/doi/s1.

**Author Contributions:** Conceptualization, C.S.; methodology, C.S. and R.D.; validation, J.S.K., C.S. and R.D.; formal analysis, J.S.K.; investigation, J.S.K., C.S. and R.D.; data curation J.S.K., R.D.; writing—original draft preparation, J.S.K. and C.S.; writing—review and editing, J.S.K., R.D. and C.S.; visualization, J.S.K.; supervision, C.S.; project administration, C.S. All authors have read and agreed to the published version of the manuscript.

**Funding:** This research received no external funding.

**Institutional Review Board Statement:** The study was conducted in accordance with the University of Pittsburgh Institutional Review Board (IRB) policy.

**Informed Consent Statement:** All interview participants provided individual consent.

**Data Availability Statement:** The raw data presented in this study are not available as per institutional IRB policy.

**Acknowledgments:** We thank Patrick Grugan for helping us in developing the initial version of the tutorial. We also thank the students and professors in the department who helped by giving constructive feedback and suggestions to make the tutorial better.

**Conflicts of Interest:** The authors declare no conflicts of interest.

## Appendix A

*Appendix A.1. Pretest*

This appendix includes the latest version of the pretest without the spaces after each question that students were provided with to answer the questions. The posttest also followed the same format with slight variation in the problems. The Supplementary Material includes the latest version of the entire tutorial, pretest, and posttest.

Pretest: Solving Laplace's Equation in Cartesian Coordinates

1. For which of the following situations is solving Laplace's equation, $\nabla^2 V = 0$, an appropriate and effective method for finding the potential $V$? (Circle correct choice/choices) Explain your reasoning.
    I. A point charge $q$ is located a distance $d$ above a grounded conducting plane. We want to find the potential at any point in space.
    II. A uniformly charged spherical volume with radius $a$ and total charge $Q$, enclosed within a concentric grounded conducting spherical shell, inner radius $b > a$, carrying charge $-Q$. We want to find the potential for $r < b$.

      III. A channel with semi-infinite conducting sides. These two sides are grounded and coincide with the planes $y = 0$ and $y = a$. They are connected in the plane $x = 0$ by a conducting sheet of width $a$ held at potential $V = V_0$ (There is a thin insulating layer at each corner so that the conductors do not touch). We want to find the potential within the channel.

2. Briefly explain the method of separation of variables applied to solve a partial differential equation such as Laplace's equation.
3. Consider the following boundary value problem: A channel consists of two parallel grounded conducting sides coinciding with the planes $y = 0$ and $y = a$. The channel is in the region $x > 0$ and is bounded at $x = 0$ with the surface set with the linear potential profile $V_0 y/a$ between the conducting sides, where $V_0$ is a constant. Find the potential within the slot.

   (a) Write down the boundary conditions for this problem.
   (b) This problem can be solved by the method of separation of variables. Which of the following choices are the orthogonal functions (in the variable $y$) which can be superposed to find the solution? In all cases, $n$ is a positive integer.

       I.   $e^{-x/a} \sin\left(\frac{n\pi y}{a}\right)$
       II.  $e^{-n\pi x/a} \sin\left(\frac{n\pi y}{a}\right)$
       III. $e^{-n\pi x/a} \cos\left(\frac{n\pi y}{a}\right)$
       IV. $e^{-n\pi x/a} \sinh\left(\frac{n\pi y}{a}\right)$
       V.  $\sinh\left(\frac{n\pi x}{a}\right) \sin\left(\frac{n\pi y}{a}\right)$

   (c) Write down the general solution for the potential $V(x,y)$ as a linear superposition of separable solutions.
   (d) According to the Fourier trick, applied to the boundary condition at $x = 0$ on the general solution, the expansion coefficients for this problem are given by

   $$C_n = \frac{2}{a}\int_0^a V(0,y) \sin\left(\frac{n\pi y}{a}\right) dy.$$

   Evaluate the integral to obtain a simplified expression for the coefficients $C_n$ [Hint: Use integration by parts].
   Note: Substitution of the expression for the coefficients $C_n$ into the general solution gives the unique solution which represents the electric potential in the region of interest for the given boundary conditions.

*Appendix A.2. Rubrics*

These rubrics followed a deduction scheme, i.e., students lose points for making errors or if they did not answer the question completely correctly. These are shown in the form of rounded off percentage deduction.

Rubrics for the latest version of the scaffolded pretest (provided in Appendix A and Supplementary Material) and posttest (only provided in Supplementary Material).

| | Pretest | |
|---|---|---|
| **Question** | **Deduction %** | **Explanation** |
| Q1 | −0% | Correct. |
| | −100% | Incorrect choice or no attempt. |
| | −67% | Circled both the correct choice and one incorrect choice. |
| Q2 | −0% | Correct. |
| | −11% | No mention of use of boundary conditions. |
| | −11% | No mention of the use of the Fourier trick to obtain the coefficients in series expansion. |
| | −11% | No mention of separation constants. |
| | −11% | No mention of separation constants summing to zero. |
| | −11% | No mention of attempt to solve by writing the solution as the product of functions in each of the variables. |
| | −11% | No mention of converting the partial differential equation to a set of ordinary differential equations. |
| | −100% | No attempt. |
| Q3 | −0% | Correct. |
| | −25% | Missing or incorrect boundary condition for $x \to \infty$. |
| | −100% | All four boundary conditions are either incorrect or missing. |
| | −75% | Three boundary conditions are either incorrect or missing. |
| | −25% | Missing or incorrect boundary condition at $x = 0$. |
| Q4 | −0% | Correct. |
| | −67% | Correct choice plus two incorrect choices. |
| | −33% | Correct choice plus one incorrect choice. |
| | −100% | Incorrect choice. |
| Q5 | −0% | Correct. |
| | −100% | Incorrect or no answer. |
| | −17% | Answer is not dimensionally correct. |
| | −67% | Miscellaneous deduction. |
| Q6 | −0% | Correct. |
| | −17% | Did not write down the expression for integrand correctly. |
| | −33% | Did not set up integration by parts. |
| | −33% | Did not correctly perform the integrals. |
| | −17% | Incorrect or no answer for the coefficients. |
| | −100% | No attempt. |
| | −8% | Minor error. |
| | −67% | Various errors or lack of progress. |

| Posttest | | |
|---|---|---|
| Question | Point Value (Deduction) | Explanation |
| Q1 | −0% | Correct. |
| | −100% | Incorrect choice or no attempt. |
| | −67% | Circled both the correct choice (II) and one incorrect choice. |
| Q2 | −0% | Correct. |
| | −11% | No mention of use of boundary conditions. |
| | −11% | No mention of the use of the Fourier trick to obtain the coefficients in series expansion. |
| | −11% | No mention of separation constants. |
| | −11% | No mention of separation constants summing to zero. |
| | −11% | No mention of attempt to solve by writing the solution as a product of functions in each of the variables. |
| | −11% | No mention of converting the partial differential equation to a set of ordinary differential equations. |
| | −100% | No attempt. |
| Q3 | −0% | Correct. |
| | −25% | Did not state: $V(x = 0, y) = 0$. |
| | −25% | Did not state: $V(x = a, y) = 0$. |
| | −25% | Did not state: $V(x, y = 0) = 0$. |
| | −25% | Did not state: $V(x, y = b) = \frac{V_o x}{a}$. |
| | −10% | Miscellaneous deduction. |
| Q4 | −0% | Correct. |
| | −67% | Correct choice plus two incorrect choices. |
| | −33% | Correct choice plus one incorrect choice. |
| | −100% | Incorrect choice. |
| Q5 | −0% | Correct. |
| | −100% | Incorrect or no answer. |
| | −17% | Answer is not dimensionally correct. |
| | −67% | Miscellaneous deduction. |
| Q6 | −0% | Correct. |
| | −17% | Did not write down the expression for integrand correctly. |
| | −33% | Did not set up integration by parts. |
| | −33% | Did not correctly perform the integrals. |
| | −17% | Incorrect or no answer for the coefficients. |
| | −100% | No attempt. |
| | −8% | Minor error. |
| | −67% | Various errors or lack of progress. |

Rubrics for the latest version of the unscaffolded pretest and posttest

- *The purpose of the rubric is to compare student performance on scaffolded versus unscaffolded tests and to offer a standardized tool that can be used by other instructors and researchers.*
- *The rubric is designed to capture whether students demonstrate the problem-solving strategies that reflect their understanding before and after engaging with the tutorial.*
- *Students are not penalized for omitting a step explicitly, as long as there is evidence elsewhere in their solution that indicates understanding of that step (this statement may override some of the deductions listed below).*
- *Each row represents a step in the problem-solving process, with steps 1–7 weighted equally for 1 point each and steps 8–12 (calculation of $C_n$) weighted equally at 0.2 each. The first column indicates the step number, while the second column describes the conceptual focus. The third column provides a descriptive rubric based on a deduction scheme. The fourth column maps each step in the unscaffolded test to the corresponding step in the scaffolded version. For Q2, in the scaffolded version, students were only asked to describe the method of separation of variable rather than use it to solve the problem (as in the unscaffolded version).*

| No. | Concept | Deduction Scheme (for Grading) |
|---|---|---|
| 1 | Diagram (optional) and boundary conditions. | −0% if diagram is missing but boundary conditions are written mathematically.<br>−0% if one of the boundary conditions is missing.<br>−50% if two of the boundary conditions are missing.<br>−50% if the diagram is drawn correctly with grounding symbols, but mathematical boundary conditions are missing.<br>−100% if three or all of the boundary conditions are missing. |
| 2 | Identify that Laplace's equation (LE) should be used for the given problem, express mathematically. | −0% if identified correctly that LE should be used and stated it in some mathematical form.<br>−50% if identified correctly that LE should be used but did not state it mathematically.<br>−100% if could not identify that solving LE is an appropriate and effective method for the given problem. |
| 3 | Identify that separation of variables (SoV) can be used to find the solution and express SoV mathematically. | −0% if SoV applied correctly, mathematically.<br>−50% if did not express SoV mathematically or did not apply correctly.<br>−100% if could not identify that SoV can be used to find the solution. |
| 4 | Write ordinary differential equations (ODEs) with correct signs. | −0% if ODEs are stated correctly with correct signs.<br>−50% if there are mistakes involving signs in ODEs.<br>−50% for other miscellaneous errors.<br>−100% if there is no attempt to write or use ODEs.<br><br>Note: No point is lost if not stated that the separation constants add up to zero. |
| 5 | Apply boundary conditions correctly and identify correct form of the separable solutions. | −0% if boundary conditions applied correctly and found correct explicit form of separable solutions.<br>−50% if there is error in applying one of the boundary conditions correctly.<br>−100% if the separable solution is completely incorrect. |
| 6 | Apply boundary condition correctly to find constants '$k_n$' (obtained from ordinary differential equation). | −0% if the correct form of '$k_n$' is obtained, with '$n$' a positive integer.<br>−50% if guessed the form of '$k_n$' but did not apply boundary condition correctly.<br>−50% if '$n$' is defined incorrectly.<br>−100% if did not find '$k_n$' at all. |
| 7 | Write general solution as a linear combination of separable solutions with coefficients. | −0% if the general solution is written correctly.<br>−50% if expansion coefficients are missing.<br>−100% if did not recognize that the general solution should be written. |
| 8 | Apply the remaining boundary condition correctly. | −0% if the correct remaining boundary condition applied correctly.<br>−50% if boundary condition is not applied correctly.<br>−100% if did not recognize that remaining boundary condition must be applied. |
| 9 | Apply the Fourier trick with correct integral and limits. | −0% if correct integral is set up.<br>−50% if incorrect limits of integral are used.<br>−100% if doesn't know how to set up integral for the Fourier trick. |
| 10 | Apply the Fourier trick with correct multiplicative function. | −0% if the correct function with appropriate index is multiplied (i.e., the index must be different from the dummy index used for writing the general solution as a linear combination of separable solutions).<br>−50% if incorrect function is multiplied (incorrect index or incorrect function).<br>−100% if the step is missing. |
| 11 | Apply orthogonality (for e.g., $\int_0^a \sin\left(\frac{n\pi y}{a}\right) \sin\left(\frac{n'\pi y}{a}\right) dy = \frac{a}{2}\delta_{nn'}$) condition correctly. | −0% if correct orthogonality relation is applied.<br>−50% if orthogonality relation is not applied correctly.<br>−100% if doesn't know that orthogonality relation should be applied. |
| 12 | Solve the integral correctly after applying the Fourier trick. | −0% if solved completely correctly.<br>−50% if there is any significant error.<br>−100% if could not identify $u, v$ in integration by parts and did not proceed to solve. |

*Appendix A.3. Previous Versions*

- Previous version 1

This version refers to the original versions of two of the questions from the pretest and posttest, as administered by I1. I2 also used this version but with the pretest and posttest swapped.

- Pretest

Q5* According to the Fourier trick, applied to the boundary condition at $x = 0$, the expansion coefficients for this problem are given by

$$C_n = \frac{2}{a} \int_0^a V(y) \sin\left(\frac{n\pi y}{a}\right) dy.$$

Evaluate the coefficients. [Hint: Use integration by parts].

Q6* Write down the solution for the potential $V(x,y)$.

- Posttest

Q5* Write down your solution for the potential as an infinite series. Then apply the boundary condition at $y = b$ to obtain the expression that you would start with to evaluate the coefficients $C_n$ by applying the Fourier trick. You are not being asked to apply the trick here, just to write down the equation obtained from the boundary condition.

- Rubric for Q5* Posttest

| Q5* | −0% | Correct. |
|---|---|---|
| | −50% | Did not write down the expression for the potential as an infinite series. |
| | −50% | Did not apply the boundary condition at $y = b$ to the infinite series. |
| | −10% | Minor error. |

Q6* After applying the Fourier trick, we obtain the following expression for the coefficients:

$$C_n = \frac{2V_0}{a^2 \sinh(n\pi b/a)} \int_0^a x \sin\left(\frac{n\pi x}{a}\right) dx.$$

Evaluate the integral to obtain a simplified expression for the coefficients $C_n$. [Hint: Use integration by parts].

- Rubric for Q6* Posttest

| Q6* | −0% | Correct. |
|---|---|---|
| | −25% | Did not identify $u$, $dv$ for integration by parts. |
| | −25% | Incorrect application of integration by parts. |
| | −25% | Integral of $v\,du$ equals zero. |
| | −25% | Incorrect answer obtained by evaluating $uv$ at the limits of integration. |

Note: The pretest and posttest administered by I3 were largely consistent with the latest versions of these tests. Q5 and Q6 appeared in reverse order in both tests for them; however, all responses were graded using the rubric from the latest versions.